\documentclass[a4paper]{article}
    \usepackage[T1]{fontenc}
    \usepackage{graphicx}
    \usepackage{epsfig}
    \usepackage{amsmath}
    \usepackage{amsfonts}
\newcommand{\be}[1]{\begin{equation}\label{#1}}
\newcommand{\ee}{\end{equation}}
\newcommand{\ba}[1]{\begin{eqnarray}\label{#1}}
\newcommand{\ea}{\end{eqnarray}}
\newcommand{\rf}[1]{(\ref{#1})}
\newcommand{\nn}{\nonumber}
\begin{document}
\fontsize{12}{12} \selectfont

\title{Dynamical dark energy from extra dimensions}

\author{V. Baukh, A. Zhuk}
\date{\begin{small}Astronomical Observatory and Department of Theoretical
Physics, \\ Odessa National University,
Odessa, Ukraine\\
bauch\_vGR@ukr.net, ai\_zhuk2@rambler.ru
\end{small}}
\maketitle

\begin{abstract}
We consider multidimensional cosmological model with a
higher-dimensional product manifold $M = \mathbb{R} \times
\mathbb{R}^{d_0} \times \mathbb{H}^{d_1}/\Gamma$ where
$\mathbb{R}^{d_0}$ is $d_0$-dimensional Ricci-flat external (our)
space and $\mathbb{H}^{d_1}/\Gamma$ is $d_1$-dimensional compact
hyperbolic internal space. M2-brane solution for this model has
the stage of accelerating expansion of the external space. We
apply this model to explain the late time acceleration of our
Universe. Recent observational data (the Hubble parameter at the
present time and the redshift when the deceleration parameter
changes its sign) fix fully all free parameters of the model. As a
result, we find that considered model has too big size of the
internal space at the present time and variation of the effective
four-dimensional fine structure constant strongly exceeds the
observational limits.
\end{abstract}

\section{Introduction}

Recent astronomical observations abundantly evidence that our
Universe underwent stages of accelerating expansion during its
evolution. There are at least two of such stages: early inflation
and late time acceleration. The latter began approximately at the
redshift $z \sim 0.35 $ (see e.g. \cite{FTH}) and continues until
now. Thus, the construction and investigation of models with
stages of acceleration is one of the main challenge of the modern
cosmology. Among such models, the models originated from
fundamental theories (e.g. string/M-theory) are of the most of
interest.

In the present paper we consider a multidimensional cosmological
model with a factorizable metric
\ba{1.1}
g &=&  -e^{2\gamma_0}d\tau \otimes d\tau + a_{0BD}^2g^{(0)} + a_1^2g^{(1)} \\
&=&  \Omega^2(-dt\otimes dt + a_0^2 g^{(0)}) + a_1^2 g^{(1)}\, ,
\nn
\ea
which is defined on the manifold with product topology
\begin{equation}
M = \mathbb{R} \times \mathbb{R}^{d_0} \times
\mathbb{H}^{d_1}/\Gamma\, ,
\end{equation}
where $\mathbb{R}^{d_0}$ is $d_0$-dimensional Ricci-flat external
(our) space with metric $g^{(0)}: R[g^{(0)}]=0$ and scale factor
$a_0$, and $\mathbb{H}^{d_1}/\Gamma$ is $d_1$-dimensional
hyperbolic (compact) internal space\footnote{Negative constant curvature
spaces are compact if they have a quotient structure: $
H^{d_i}/\Gamma_i$, where $H^{d_i}$ and $\Gamma_i$ are hyperbolic
spaces and their discrete isometry group, respectively.} with metric $g^{(1)}:
R[g^{(1)}]=-d_1(d_1-1)$ and scale factor $a_1$. Both $a_0$ and
$a_1$ depend only on time. First line in \rf{1.1} is a metric in
the Brans-Dicke frame in the harmonic time gauge where
$e^{\gamma_0}=a_{0BD}^{d_0}a_1^{d_1}$ \cite{IMZ}. Second line in
\rf{1.1} is a metric in the Einstein frame in the synchronous time
gauge. The scale factors $a_{0BD}$ of the external space in the Brans-Dicke frame is connected with the scale factor $a_0$ in the Einstein frame as follows: $a_0=\Omega^{-1}a_{0BD}$, where conformal factor $\Omega=a_1^{-\frac{d_1}{d_0-1}}$.
Hereafter we consider 3-dimensional external space: $d_0=3$.
Harmonic time $\tau$ is related to synchronous time $t$ as
$dt=f(\tau)d\tau$, where $f(\tau)=\Omega^{-1}a_{0BD}^{3}a_1^{d_1}=
a_{0BD}^{3}a_1^{3d_1/2}=a_0^{3}$ \cite{BaukhZhuk}.

With the standard York-Gibbons-Hawking boundary term $S_{YGH}$, an
action for considered model reads
\be{1.2}
S = \frac{1}{2\kappa^2}\int\limits_{M}d^Dx\sqrt{|g|}R[g]+S_{YGH}\,
,
\ee
where $D=1+d_0+d_1 =4+d_1$ is a total number of dimensions and
$\kappa_D$ is a $D$-dimensional gravitational constant.

Substituting the metric \rf{1.1} into this action and minimizing
obtained Lagrangian with respect to the scale factors, we get the
the following solutions\footnote{The general method for this kind
of models was elaborated in papers \cite{Zhuk1,BZnegative}.} of
the equations of motion (in the harmonic time gauge):
\be{1.3}
a_0(\tau)=A_1^{\frac{d_1+2}{6}}
\left(\sqrt{\frac{2\varepsilon}{|R_1|}}\right)^{\frac{d_1}{2(d_1-1)}}
\frac{\exp\left(-\sqrt{\frac{d_1+2}{12(d_1-1)}{2\varepsilon}}\; \tau\right)}
{\sinh^{d_1/[2(d_1-1]}(-\sqrt{\frac{d_1-1}{d_1}
2\varepsilon}\; \tau)}\, ,
\ee
and
\be{1.4}
a_1(\tau)=A_1\left(\sqrt{\frac{2\varepsilon}{|R_1|}}\right)^{\frac{1}{d_1-1}}
\frac{\exp\left(-\sqrt{\frac{3}{(d_1-1)(d_1+2)}2\varepsilon}{\; \tau}\right)}
{\sinh^{1/(d_1-1)}(-\sqrt{\frac{d_1-1}{d_1}
2\varepsilon}\; \tau)}\, ,
\ee
where  $A_1$ and $\varepsilon$ are the constants of integration.
The function $f(\tau)$ can be easily obtained from Eq. \rf{1.3}
via expression $f(\tau) = a_0^3(\tau)$.

We should note, that solutions \rf{1.3}  and \rf{1.4} for the
metric \rf{1.1} is a particular case of so called S$p$-branes with
$(p+1)-$dimensional Ricci-flat external space. In the case $d_0=3$
we obtain $p=2$. Therefore, if underlying model is
$(D=11)-$dimensional M-theory, we arrive at M2-branes where the
number of internal dimensions is equal to 7. It is well known that
such models with hyperbolic internal space undergo the stage of
accelerating expansions (see e.g. \cite{BaukhZhuk} and references
therein). However, the parameters of the model were not connected
with observational data. So, in the present paper we want to use
the modern cosmological data (the present day value for the Hubble
parameter and the redshift when our external space transits from
deceleration to acceleration) to fix all arbitrary parameters of
the considered model and obtain corresponding dynamical behavior
for the scale factors, the Hubble parameter, the deceleration
parameter and the fine structure "constant".

\section{Dynamical behavior of the model}

In this section, besides the external $a_0$ and internal $a_1$
scale factors described by Eqs. \rf{1.3} and \rf{1.4}, we consider
also the Hubble parameter for each of the factor spaces
\be{2.1}
H_0 = \frac{1}{a_0}\frac{da_0}{dt}=\frac{1}{a_0f(\tau )
}\frac{da_0}{d\tau}=
-\frac{\sqrt{2\varepsilon}}{f(\tau)}\left(\sqrt{\frac{d_1+2}{12(d_1-1)}}+
\sqrt{\frac{d_1}{4(d_1-1)}}\coth\left(\sqrt{\frac{d_1-1}{d_1}2\varepsilon}\;
\tau\right)\right)\, ,
\ee
\be{2.2}
H_1=\frac{1}{a_1}\frac{da_1}{dt}=\frac{1}{a_1f(\tau )
}\frac{da_1}{d\tau}=-\frac{\sqrt{2\varepsilon}}{f(\tau)}\sqrt{\frac{3}{(d_1-1)(d_1+2)}}
\left(1+\sqrt{\frac{d_1+2}{3d_1}}
\coth\left(\sqrt{\frac{d_1-1}{d_1}2\varepsilon}\;
\tau\right)\right)\, ,
\ee
the external space deceleration parameter
\ba{2.3} q_0&=&-\frac{d^2a_0}{dt^2}\frac{1}{H^2_0a_0}=
-\frac{1}{f(\tau)}\frac{d}{d\tau}\left(\frac{1}{f(\tau)}\frac{da_0}{d\tau}\right)\frac{1}{H^2_0a_0}\nn
\\
&=& -2\sinh^{-2}(\sqrt{\frac{d_1-1}{d_1}2\varepsilon}\; \tau)
\left[\sqrt{\frac{d_1+2}{3(d_1-1)}}+\sqrt{\frac{d_1}{d_1-1}}
\coth\left(\sqrt{\frac{d_1-1}{d_1}2\varepsilon}\;
\tau\right)\right]^{-2} +2
\ea
and the variation of the fine structure constant\footnote{It is
well known that the internal space dynamics results in the
variation of the fundamental constants such as the fine structure
constant (see, e.g., Refs. \cite{Uzan,Kub,CV,GSZ}). For example,
the effective four-dimensional fine-structure constant is
inversely proportional to the volume of the internal space:
$\alpha  \sim V^{-1}_{1} \sim a_{1}^{-d_{1}}$.} (as a function of
redshift $z$)
\be{2.4}
\Delta \alpha=\frac{\alpha(z) -\alpha(0)}{\alpha(0)}=
\frac{a_1^{2d_1/(d_0-1)}(0)}{a_1^{2d_1/(d_0-1)}(z)}-1\, .
\ee
We
assume also that the solution \rf{1.3},\rf{1.4} describes the
M2-brane, that is $d_1=7$.

According to the recent observational data (see e.g.
\cite{FTH,WMAP5}), the present acceleration stage began at
redshift $z\approx 0.35$ and the Hubble parameter now is $H_0
(t_p) \equiv H_p \approx 72\, \mbox{km}/\mbox{sec}/\mbox{Mpc} =
2.33 \times 10^{-18} \mbox{sec}^{-1}$. Hereafter, the letter $p$
denotes the present day values. Additionally, at the present time
the value of the external space scale factor can be estimated as
$a_0(t_p)\approx cH_p^{-1} \approx 1.29\times 10^{28}$cm.  We
shall use these observational conditions to fix the free
parameters of the model $A_1$ and $\varepsilon$ (the constants of
integration) and to define the present time\footnote{It is obvious
that our model cannot pretend to describe the full history of the
Universe. We try to apply this model to explain the late time
acceleration of the Universe which starts at the redshift
$z\approx 0.35$. Before this time, the Universe evolution is
described by the standard Big Bang cosmology. Therefore, in our
model $t=0$ corresponds to $z=0.35$ (i.e. $q_0(z=0.35)=0$) and
$t_p$ is the time from this moment to the present day.} $t_p$.
Observational data also show that for different redshifts the fine
structure constant variation does not exceed $10^{-5}$: $|\Delta
\alpha|<10^{-5}$ \cite{Uzan}.

Below, all quantities are measured in the Hubble units. For
example, the scale factors are measured in $cH_p^{-1}$ and
synchronous time $t$ is measured in $H_p^{-1}$. Therefore,
$a_0(t_p)=1$ and $H_p=1$.

To fix all free parameters of the model, we use the following
logic chain. First, from the equation $q_0(\tau)=0$ we obtain the
harmonic time $\tau_{in}$ of the beginning of the stage of
acceleration. We find that this equation has two roots which
describe the beginning and end of the acceleration. Second, we
define the constant of integration $A_1$ from the equation
$z=0.35=1/a_0(\tau_{in})-1$ where we use the condition that
acceleration starts at $z=0.35$ and that $a_0(\tau_p)=1$. Third,
we find the present harmonic time $\tau_p$ from the condition
$a_0(\tau_p)=1$. It is worth of noting that $\tau_{in}, \tau_{p}$
and $A_1$ are the functions of $\varepsilon$. To fix this
parameter, we can use the condition $H_0(\tau_p)=1$. Finally, to
find the value of the present synchronous time, we use the
equation $t_p=\int_{\tau_{in}}^{\tau_p}f(\tau)d\tau$ where
$f(\tau)=a_0^3(\tau)$. In the case $d_1=7$, direct calculations
give for the constants of integration $A_1=1.23468$ and
$\varepsilon=1.53097$. It results in $t_{p}=0.296\sim
4\mbox{Gyr},\quad q_0(t_{p})=-0.960572$ and for the internal space
$a_1(t_{p})=1.24319, \quad H_1(t_{p})=0.0500333$

Dynamical behavior of the considered model is depicted in figures
1-3. Fig.1 shows the dynamics of the external space scale factor
$a_0(t)$ (left panel) and the internal space scale factor
$a_1(t)$( right panel). Here, $t=0.296$ is the present time, and
$t=0$ and $t=1.28$ correspond to the beginning and end of the
stage of acceleration, respectively. It follows that the internal
space is the same order of magnitude as the external one at the
present time. However, for the standard Kaluza-Klein models there
is the experimental restriction on the size of extra dimensions:
$l_{extra}\leq 10^{-17}$cm. That is $a_0/a_1 \geq 10^{45}$.
Obviously, our model does not satisfy this condition. One of the
possible way to avoid this problem consists in proposal that the
Standard Model matter is localized on a brane. In this case the
extra dimensions can be much bigger that $10^{-17}$cm (even an
infinite). However, such model requires the generalization of our
metric \rf{1.1} to the non-factorizable case and this
investigation is out of the scope of the present paper.

We plot in Fig.2 the evolution of the Hubble parameters $H_0(t)$
(left panel) and $H_1(t)$ (right panel). We can see that their
values are comparable with each other. Thus, the internal space is
rather dynamical and this fact is the main reason of too large
variations of the fine structure constant (see Fig. 3).

We present in Fig. 3 the evolution of the deceleration parameter
$q_0(t)$ (left panel) and the variation of the fine structure
constant $\Delta \alpha (t)$ (right panel). Left picture clearly
shows that the acceleration stage has the final period for the
considered model. It starts at $t=0$ and finishes at $t=1.28$. The
right picture demonstrates that $\Delta \alpha$ does not satisfy
the observable restrictions $|\Delta \alpha|<10^{-5}$. There is
the only very narrow region in the vicinity of $z=0.13$ (or equivalently $t=0.17$ in synchronous time) where
$\Delta \alpha$ changes its sign. However, it is the exceptional
region but restriction $|\Delta \alpha|<10^{-5}$ works for very
large diapason of redshifts $z$ \cite{Uzan}.

\begin{figure*}[htbp]
\centerline{ \includegraphics[width=3in,height=2.5in]{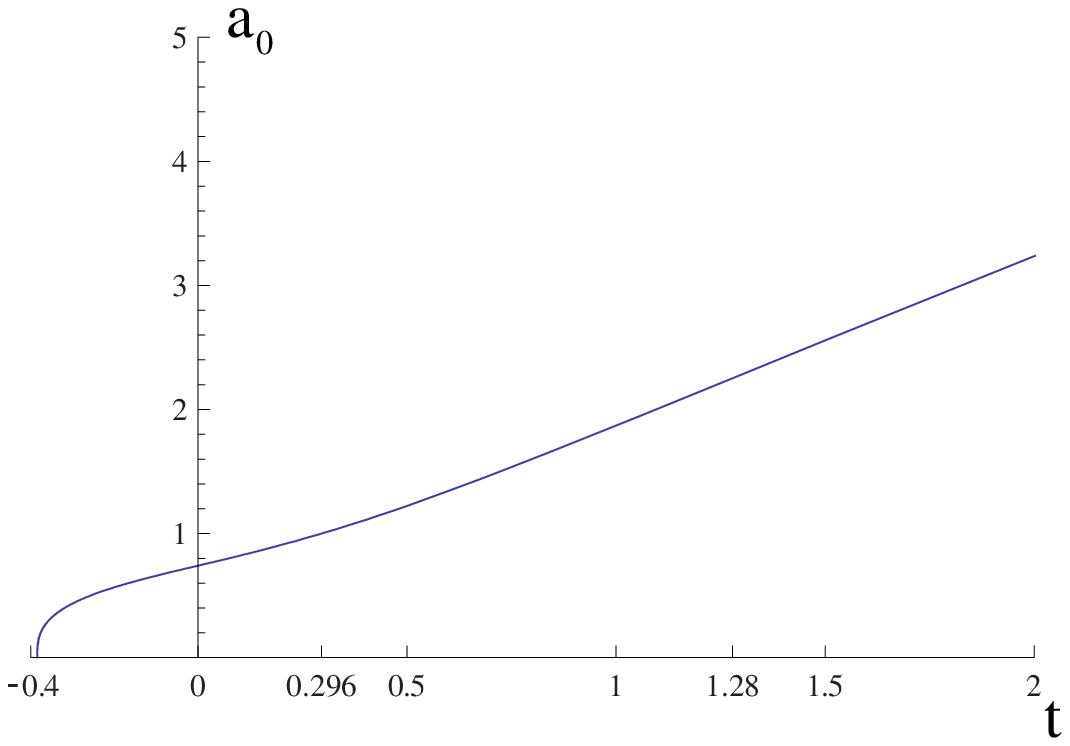}
\includegraphics[width=3in,height=2.5in]{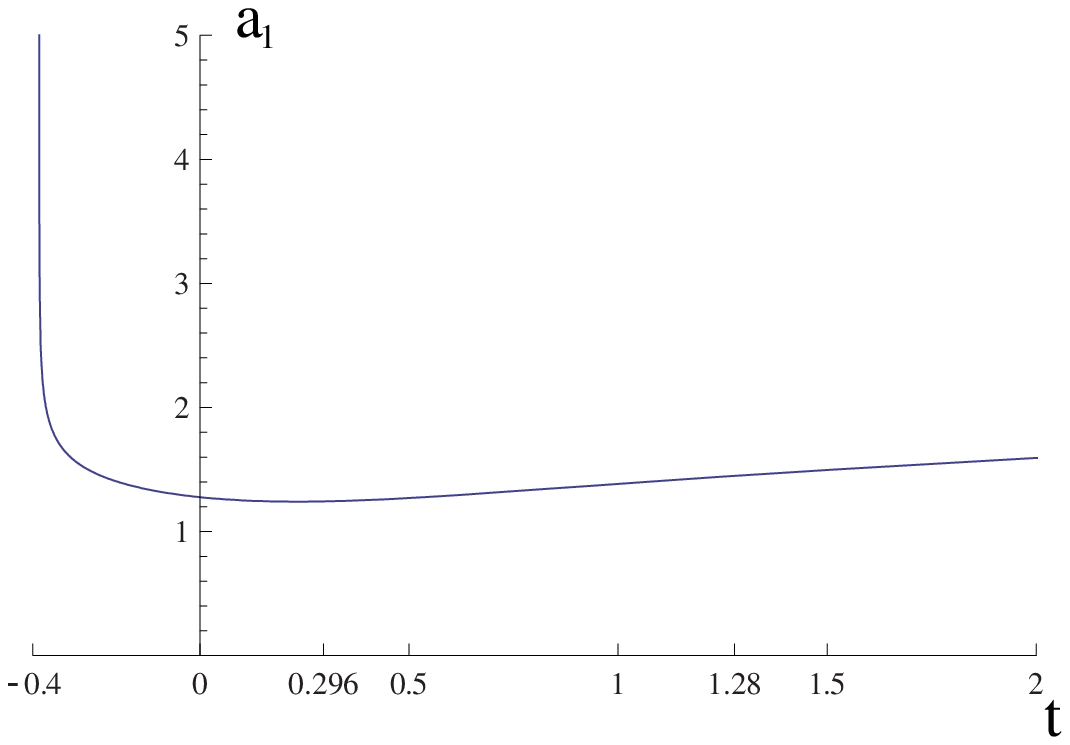}}
\caption{The scale factors of the external space (left panel) and
internal space (right panel) versus synchronous time.}
\end{figure*}

\begin{figure}[htbp]
\centerline{ \includegraphics[width=3in,height=2.5in]{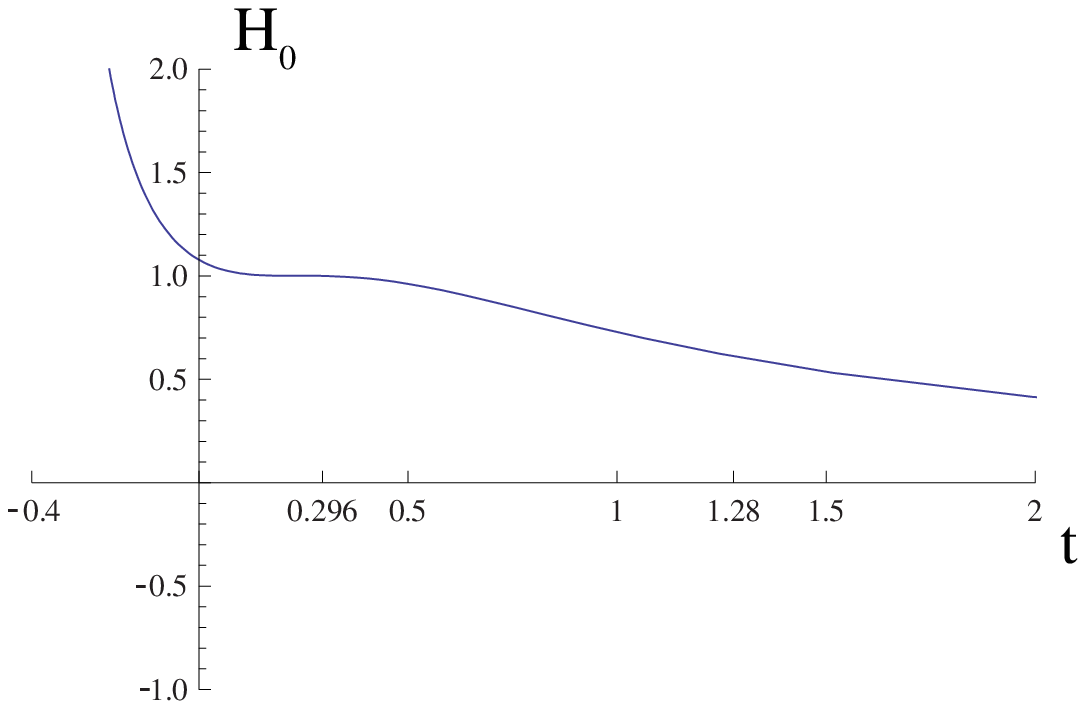}
\includegraphics[width=3in,height=2.5in]{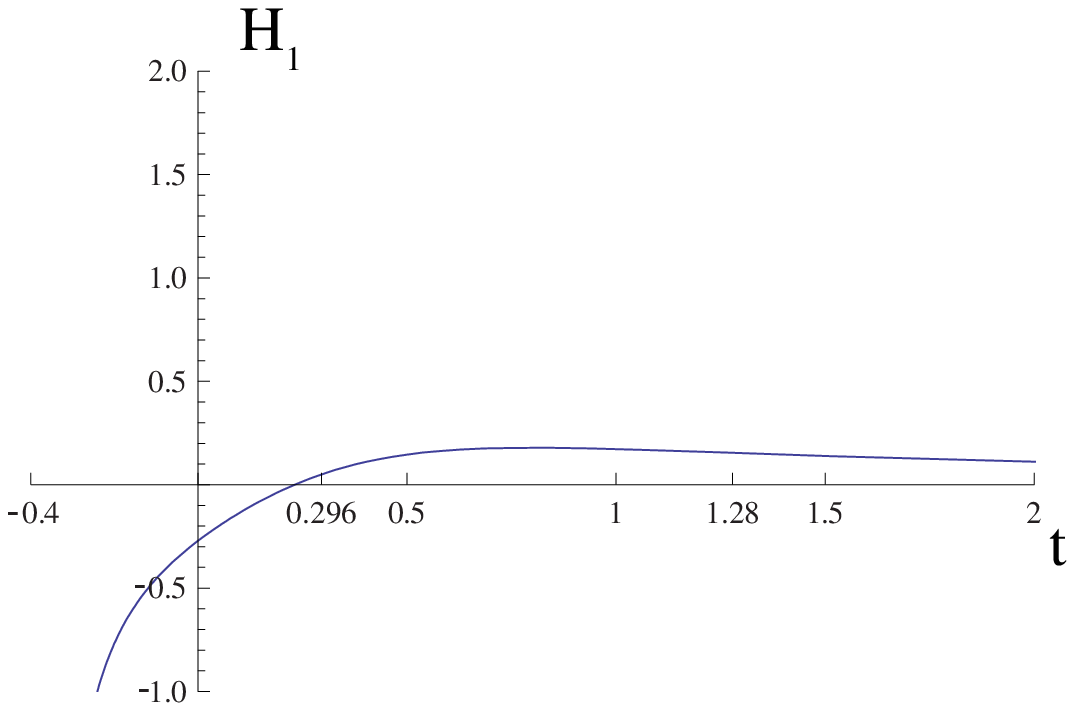}}
\caption{The Hubble parameters of the external space (left panel)
and internal space (right panel) versus synchronous time.}
\end{figure}

\begin{figure}[htbp]
\centerline{ \includegraphics[width=3in,height=2.5in]{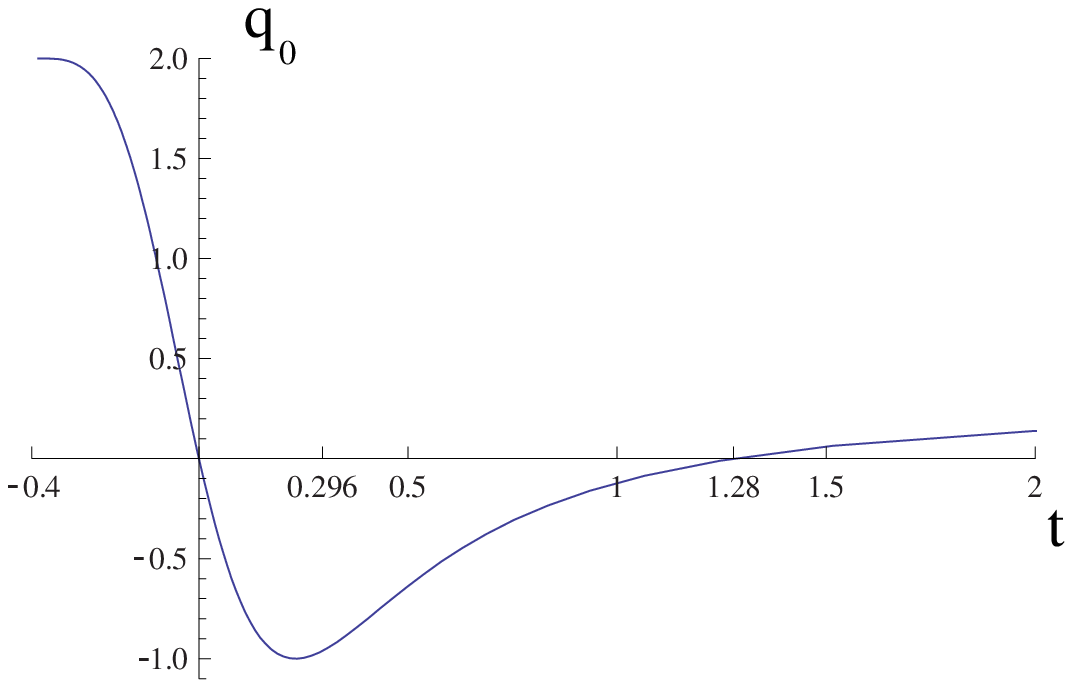}
\includegraphics[width=3in,height=2.5in]{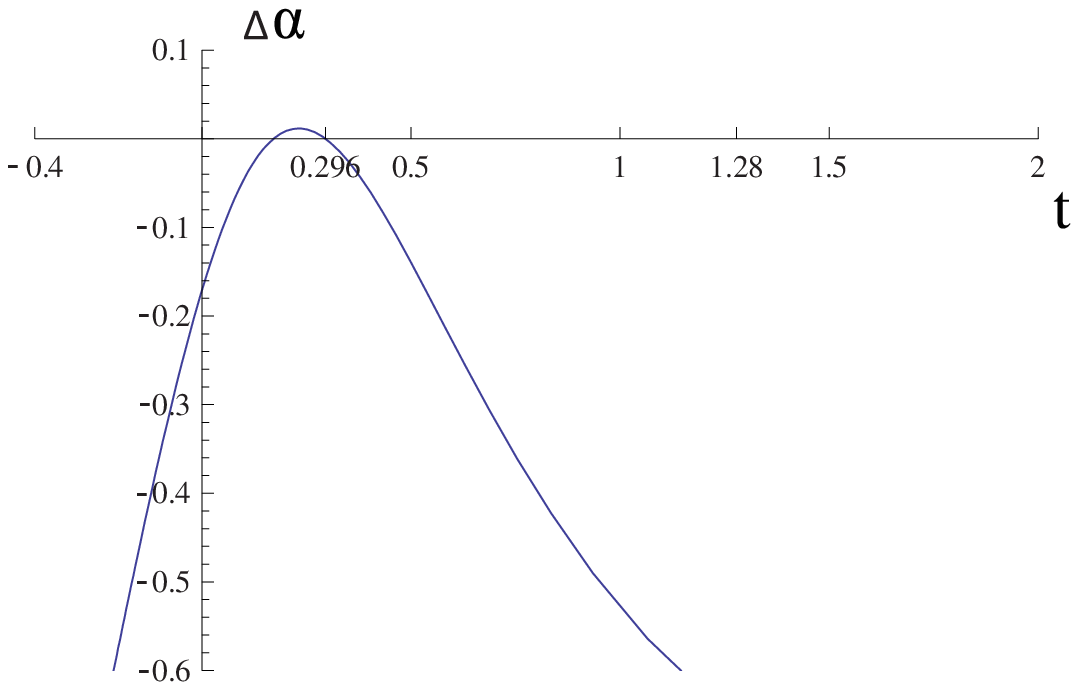}}
\caption{The deceleration parameter of the external space (left
panel) and variation of the fine structure constant (right panel)
versus synchronous time.}
\end{figure}

\section{Conclusions}

In the present paper we investigate multidimensional cosmological
model with Ricci-flat external space and compact hyperbolic
internal space. Such pure gravitational model has the exact
solution with the stage of accelerating expansion for our external
space (so called Sp-brane solution). This solution depends on a
number of free parameters. It is remarkable that observable
cosmological parameters (such as the Hubble parameter, the
parameter of deceleration) gives a possibility to fix all these
free parameters and completely determinate this model. We perform
this analysis in the case of M2-brane with $d_1=7$ extra
dimensions. For obtained parameters, we describe the dynamical
behavior of the considered model. It is shown that our external
space really has the finite stage of the accelerating expansion
starting at $z=0.35$ and lasting until now. However, this model
has two significant drawbacks. On the one hand, the internal space
is too big with respect to the standard Kaluza-Klein restrictions
$a_{internal}\leq 10^{-17}$cm and, on the other hand, this space is not
sufficiently constant to satisfy the observable limits on the fine
structure constant variations. Thus, these drawbacks rule out this
model from the scope of viable theories\footnote{ In papers \cite{MB}, it is investigated the effects of matter overdensities on the
time and space variations of alpha. It is shown that a far slower evolution of alpha will be found in
virialised regions (such as the Earth or the Solar System) than in the
cosmological background. Thus, this effect gives a possible way to alleviate our conclusions concerning incompatibleness of the considered model 
with the observations on variations of alpha. However, the problem of too big size of the extra dimensions still remains.}.

\section*{Acknowledgements}
\indent \indent This work was supported in part by the
"Cosmomicrophysics" programme of the Physics and Astronomy
Division of the National Academy of Sciences of Ukraine. We thank David Mota for valuable correspondence and calling our attention to their papers \cite{MB} which shed light upon a possible way to alleviate our conclusions concerning incompatibleness of the considered model
with the observations on variations of alpha.

\end{document}